\documentclass[12pt]{article}
\usepackage{axodraw}
\usepackage{epsfig}
\def\e{{\rm e}}
\newcommand{\be}{\begin{equation}}
\newcommand{\ee}{\end{equation}}
\newcommand{\bea}{\begin{eqnarray}}
\newcommand{\eea}{\end{eqnarray}}

\newcommand{\gm}{\gamma}
\newcommand{\Gm}{\Gamma}

\newcommand{\ep}{\epsilon}

\newcommand{\dd}{\mbox{d}}

\newcommand{\nn}{\nonumber}

\newcommand{\Li}[2]{{\mbox{Li}}_{#1}\left(#2\right)}

\begin{document}
\parindent=1.5pc

\begin{titlepage}
\leftline{DESY~01-190\hfill ISSN 0418-9833}
\leftline{hep-ph/0111160}
\leftline{November 2001}
\bigskip
\begin{center}
{{\bf
Analytical Result for Dimensionally Regularized Massive
On-Shell Planar Double Box
} \\
\vglue 5pt
\vglue 1.0cm
{ {\large V.A. Smirnov\footnote{
Permanent address: Nuclear Physics Institute of Moscow State University,
Moscow 119899, Russia.
E-mail: smirnov@theory.sinp.msu.ru}
} }\\
\baselineskip=14pt
\vspace{2mm}
{\normalsize II. Institut f\"ur Theoretische Physik, Universit\"at Hamburg,}\\
{\normalsize Luruper Chaussee 149, 22761 Hamburg, Germany}
\vglue 0.8cm
{Abstract}}
\end{center}
\vglue 0.3cm
{\rightskip=3pc
 \leftskip=3pc
\noindent The dimensionally regularized master planar double box
{}Feynman diagram  with four massive and three massless lines,
powers of propagators equal to one, all four legs
on the mass shell, i.e. with $p_i^2=m^2,\;i=1,2,3,4$, is
analytically evaluated for general
values of $m^2$ and the Mandelstam variables $s$ and $t$. An explicit
result is expressed in terms of polylogarithms, up to the
third order,  depending on special combinations of $m^2,s$ and~$t$.
\vglue 0.8cm}
\end{titlepage}


{\bf 1.}
Systematical analytical evaluation of two-loop Feynman diagrams with
four external lines within dimensional regularization \cite{dimreg}
began two and a half years ago.
In the pure massless case with all end-points on-shell, i.e.
$p_i^2=0,\;i=1,2,3,4$, the problem of analytical evaluation
of two-loop four-point diagrams in expansion in $\ep=(4-d)/2$,
where $d$ is the space-time dimension, has been completely solved
in \cite{K1,SV,Tausk,AGO,ATT}.
Any such diagram can be expressed, in Laurent expansion in $\ep$ up to
a finite part, through polylogarithms and generalized polylogarithms
up to the fourth order, depending on the ratio of the Mandelstam
variables $s$ and $t$.
The corresponding analytical algorithms have been successfully
applied to the evaluation of two-loop virtual
corrections to various scattering processes \cite{appl} in the
zero-mass approximation.

If in addition to the Mandelstam variables $s$ and $t$
there is one more massive parameter, the four-point diagrams become
much more complicated. However, in the case of one leg off-shell relevant
to the process $e^+e^-\to 3$jets (see, e.g., \cite{3jets}), the problem
of the evaluation has been solved: master integrals have been evaluated
\cite{S2,GR2}, either by use of Mellin--Barnes representation or
the method of differential equations \cite{DE},
and a reduction procedure has been developed \cite{GR2}.
(See \cite{Moch} where the present status of NNL0 calculations
of the process $e^+e^-\to 3$jets is characterized.)

The purpose of this paper is to turn attention to on-shell four-point diagrams
with a non-zero internal mass and analytically evaluate the scalar
double box diagram shown in Fig.~1.
\vspace{0.5cm}
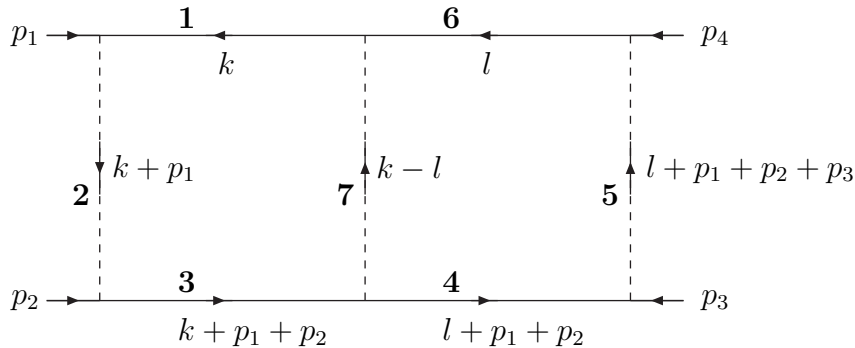
\begin {figure} [htbp]
\begin{picture}(400,100)
\ArrowLine(80,0)(100,0)
\ArrowLine(80,100)(100,100)
\ArrowLine(320,0)(300,0)
\ArrowLine(320,100)(300,100)
\Line(80,0)(320,0)
\Line(80,100)(320,100)
\DashLine(100,0)(100,100){3}
\DashLine(200,0)(200,100){3}
\DashLine(300,0)(300,100){3}
\ArrowLine(150,100)(140,100)
\ArrowLine(250,100)(240,100)
\ArrowLine(140,0)(150,0)
\ArrowLine(240,0)(250,0)
\ArrowLine(100,60)(100,40)
\ArrowLine(300,40)(300,60)
\ArrowLine(200,40)(200,60)
\Text(105,50)[l]{$ k+p_1$} %
\Text(205,50)[l]{$ k- l$}
\Text(307,50)[l]{$ l+p_1+p_2+p_3$}%
\Text(130,-12)[l]{$ k+p_1+p_2$} %
\Text(230,-12)[l]{$ l+p_1+p_2$} %
\Text(145,90)[l]{$ k$} %
\Text(245,90)[l]{$ l$}  %
\Text(230,107)[l]{\bf 6} %
\Text(130,107)[l]{\bf 1} %
\Text(230,7)[l]{\bf 4}  %
\Text(130,7)[l]{\bf 3}
\Text(90,40)[l]{\bf 2}  %
\Text(190,40)[l]{\bf 7} %
\Text(290,40)[l]{\bf 5} %
\Text(67,0)[l]{$p_2$}
\Text(67,100)[l]{$p_1$}
\Text(328,0)[l]{$p_3$}
\Text(328,100)[l]{$p_4$}
\end{picture}
\vspace*{2mm}\\
\caption{Planar double box diagram. Solid and dashed lines denote
massive and massless propagators, respectively.}
\end{figure}
%
The calculational experience, in particular obtained in the above
mentioned works on the evaluation of four-point diagrams,
tell us that if such master integrals can be evaluated, the problem
can be also completely solved, after evaluating other master integrals
and constructing a recursive procedure that expresses any given Feynman
integral with general numerators and integer powers of propagators
through the master integrals. Therefore this explicit analytical result
can be considered as a kind of existence theorem, in the sense that it
strongly indicates the possibility to
analytically compute various scattering processes in two loops
without putting masses to zero.

To arrive at this result we derive in the next section an appropriate
Mellin--Barnes (MB) representation for the general planar  double box within
dimensional regularization.
Then we turn to the master double box and
use a standard procedure of taking residues and shifting
contours to resolve the structure of singularities in
the parameter of dimensional regularization, $\ep$.
The initial MB integral is eventually decomposed into seven pieces
where expansion of the integrand in $\ep$ becomes possible. After evaluating
these expanded integrals we obtain an explicit analytical result
expressed in terms of polylogarithms, up to the
third order,  depending on special combinations of $m^2,s$ and~$t$.
In the last section,  leading order terms of the asymptotic expansion
of the considered diagram in the limit of small $m$ obtained by the
strategy of expansion by regions \cite{BS,SR,Sb} are presented. They serve as
a crucial check of the result obtained. We conclude with a discussion
of the results of the paper and open problems.

{\bf 2.}
Let us consider the general on-shell double box diagram of Fig.~1, i.e.
with general irreducible numerator and powers of propagators.
We choose  this irreducible numerator and
the routing of the external momenta as in \cite{ATT}. For convenience,
we consider the factor with $(k+p_1+p_2+p_3)^2$ corresponding
to the irreducible numerator
as an extra propagator but, really, we are interested only in the
non-positive integer values of $a_8$.
This general double box Feynman integral takes the form
\bea
B(a_1,\ldots,a_8;s,t,m^2;\ep) &=&
\int\int \frac{\dd^dk \, \dd^dl}{(k^2-m^2)^{a_1}[(k+p_1)^2)]^{a_2}
[(k+p_1+p_2)^2-m^2]^{a_3}}
\nn \\ && \hspace*{-50mm}
\times \frac{[(k+p_1+p_2+p_3)^2]^{-a_8}}{
[(l+p_1+p_2)^2-m^2)]^{a_4}[(l+p_1+p_2+p_3)^2]^{a_5}
(l^2-m^2)^{a_6} [(k-l)^2]^{a_7} }
\, ,
\label{2box}
\eea
where $s=(p_1+p_2)^2, \;  t=(p_2+p_3)^2$, and
$k$ and $l$ are respectively loop momenta of the left and the right box.
Usual prescriptions $k^2=k^2+i 0, \; s=s+i 0$, etc. are implied.

To resolve the singularity structure of Feynman integrals in $\ep$
it is very useful to apply the MB representation
\be
\frac{1}{(X+Y)^{\nu}} = \frac{1}{\Gm(\nu)}
\frac{1}{2\pi i}\int_{-i \infty}^{+i \infty} \dd z
\frac{Y^z}{X^{\nu+z}} \Gm(\nu+z) \Gm(-z) \;,
\label{MB}
\ee
that makes it
possible to  replace sums of terms raised to some power by their
products in some powers, at the cost of introducing an extra
integration.
In \cite{K1,Tausk,S2}
MB integrations were introduced directly in alpha/Feynman
parametric integrals.
It turns out more convenient
to follow (as in \cite{ATT}) the strategy of \cite{UD} and introduce,
in a suitable way,
MB integrations, first, after integration over one of the loop momenta, $l$,
and complete this procedure after integration
over the second loop momentum, $k$. In fact, the procedure of \cite{ATT}
is straightforwardly generalized by introducing two extra MB integrations
when separating terms with $m^2$ after each of the integrations over
the loop momenta, and after appropriate changes of variables
leads to the following sixfold MB representation of
(\ref{2box}):
\bea
B(a_1,\ldots,a_8;s,t,m^2;\ep)
 &=&
\frac{\left(i\pi^{d/2} \right)^2 (-1)^a}{
\prod_{j=2,4,5,6,7}\Gm(a_j) \Gm(4-a_{4567}-2\ep)(-s)^{a-4+2\ep}}
\nn \\ &&  \hspace*{-50mm}\times
\frac{1}{(2\pi i)^6} \int_{-i\infty}^{+i\infty}
\dd w \prod_{j=1}^5 \dd z_j
\left(\frac{m^2}{-s} \right)^{z_1+z_5}
\left(\frac{t}{s} \right)^{w}
\frac{\Gm(a_2 + w) \Gm(-w) \Gm(z_2 + z_4) \Gm(z_3 + z_4)}
{\Gm(a_1 + z_3 + z_4) \Gm(a_3 + z_2 + z_4)}
\nn \\ &&  \hspace*{-50mm}\times
\frac{  \Gm(4 - a_{13} - 2a_{28} - 2 \ep + z_2 + z_3)
  \Gm(a_{1238} - 2 + \ep + z_4 + z_5)\Gm(a_7 + w - z_4)
  }
{\Gm(4 - a_{46} - 2 a_{57} - 2 \ep - 2 w - 2 z_1 - z_2 - z_3)}
\nn \\ &&  \hspace*{-50mm}\times
\frac{\Gm(a_{4567} - 2 + \ep + w + z_1 - z_4)
  \Gm(a_8 - z_2 - z_3 - z_4) \Gm(-w - z_2 - z_3 - z_4)}
{\Gm(4 - a_{1238} - 2 \ep + w - z_4)\Gm(a_8 - w - z_2 - z_3 - z_4) }
\nn \\ &&  \hspace*{-50mm}\times
\frac{\Gm(a_5 + w + z_2 + z_3 + z_4)
  \Gm(2 - a_{567} - \ep - w - z_1 - z_2) }
  {\Gm(4 - a_{13} - 2 a_{28} - 2 \ep + z_2 + z_3 - 2 z_5)}
  \nn \\ &&  \hspace*{-50mm}\times
\Gm(2 - a_{457} - \ep - w - z_1 - z_3)
  \Gm(2 - a_{128} - \ep + z_2 - z_5) \Gm(2 - a_{238} - \ep + z_3 - z_5)
  \nn \\ &&  \hspace*{-50mm}\times
  \Gm(4 - a_{46} - 2 a_{57} - 2 \ep - 2 w - z_2 - z_3) \Gm(-z_1)\Gm(-z_5)
\, ,
\label{6MB}
\eea
where $a_{4567}=a_4+a_5+a_6+a_7, a_{13}=a_1+a_3$, etc., and
integration contours are chosen in the standard way.

In the case of the master double box, we set $a_i=1$ for $i=1,2,\ldots,7$
and $a_8=0$ and obtain
\bea
B^{(0)}(s,t,m^2;\ep)\equiv B(1,\ldots,1,0;s,t,m^2;\ep) &&
\nn \\ &&  \hspace*{-75mm}
= -\frac{\left(i\pi^{d/2} \right)^2}{
\Gm(-2\ep)(-s)^{3+2\ep}}
\frac{1}{(2\pi i)^6} \int_{-i\infty}^{+i\infty}
\dd w \prod_{j=1}^5 \dd z_j
\left(\frac{m^2}{-s} \right)^{z_1+z_5}
\left(\frac{t}{s} \right)^{w}
\frac{ \Gm(1 + w)\Gm(-w) }{\Gm(1 - 2 \ep + w - z_4) }
\nn \\ &&  \hspace*{-75mm}\times
\frac{ \Gm(2 + \ep + w + z_1 - z_4)\Gm(-1 - \ep - w - z_1 - z_2)
\Gm(-1 - \ep - w - z_1 - z_3) \Gm(-z_1)}
{\Gm(1 + z_2 + z_4) \Gm(1 + z_3 + z_4)
\Gm(-2 \ep + z_2 + z_3 - 2 z_5)}
\nn \\ &&  \hspace*{-75mm}\times
\frac{\Gm(-\ep + z_2 - z_5) \Gm(-\ep + z_3 - z_5)
\Gm(1 + \ep + z_4 + z_5)\Gm(-z_5) \Gm(-2 \ep + z_2 + z_3)  }
{\Gm(-2 - 2 \ep - 2 w - 2 z_1 - z_2 - z_3)}
\nn \\ &&  \hspace*{-75mm}\times
\Gm(-2 - 2 \ep - 2 w - z_2 - z_3)
\Gm(1 + w + z_2 + z_3 + z_4)   \Gm(z_2 + z_4) \Gm(z_3 + z_4)
\nn \\ &&  \hspace*{-75mm}\times
\Gm(1 + w - z_4)\Gm(-z_2 - z_3 - z_4)
\, .
\label{6MB0}
\eea
Observe that, because of the presence of the factor $\Gm(-2\ep)$
in the denominator, we are forced to take some residue
in order to arrive at a non-zero result at $\ep=0$,
so that the integral is effectively fivefold.

The resolution of singularities in $\ep$ is performed also in the
standard way (see \cite{K1,Tausk,ATT,S2}) and reduces to shifting
contours and taking residues. The goal of this procedure is to decompose
a given integral into pieces where the Laurent expansion $\ep$
of the integrand becomes possible.
This is how such procedure can be performed for (\ref{6MB0}):

\begin{enumerate}
\item
Take minus residue at $z_3=-2 - 2 \ep - 2 w - z_2$, then
minus residue at $w=-1-2\ep$, then residue at $z_4=0$,
then residue at $z_2=0$, expand in a Laurent series in $\ep$
up to a finite part.
Let us denote the resulting integral over $z_1$ and $z_5$ by $B_1$.
\item
Take minus residue at $z_3=-2 - 2 \ep - 2 w - z_2$, then
minus residue at $w=-1-2\ep$, then residue at $z_4=0$,
and change the nature of the first pole of $\Gm(z_2)$ (choose
a contour from the opposite side, i.e. the pole $z_2$ will be now
to the right of the contour), then expand in $\ep$.
Denote this integral over $z_1$, $z_2$ and
$z_5$ by $B_2$.
\item
Take minus residue at $z_3=-2 - 2 \ep - 2 w - z_2$, then
minus residue at $w=-1-2\ep$, then change the nature
of the first pole of $\Gm(z_4)$,
then take a residue at $z_2=-z_4$, then take a residue at $z_4=-\ep$
and expand in $\ep$.
This resulting integral over $z_1$ and $z_5$ is denoted by $B_3$.
\item
Take minus residue at $z_3=-2 - 2 \ep - 2 w - z_2$, then
minus residue at $w=-1-2\ep$, then change the nature
of the first pole of $\Gm(z_4)$,
then take a residue at $z_2=-z_4$, then
change the nature of the first pole of $\Gm(2(\ep+z_4))$ and
expand in $\ep$.
The resulting integral over $z_1,z_4$ and $z_5$ is denoted by $B_4$.
\item
Take minus residue at $z_3=-2 - 2 \ep - 2 w - z_2$, then
minus residue at $w=-1-2\ep$, then change the nature of the
first pole of $\Gm(z_4)$,
then change the nature of the first pole of $\Gm(z_2+z_4)$ and
expand in $\ep$.
The resulting integral over $z_1,z_2,z_4$ and $z_5$ is denoted by $B_5$.
\item
Take minus residue at $z_3=-2 - 2 \ep - 2 w - z_2$, then
change the nature of the first pole of $\Gm(-2(1+2\ep+w))$, then
take minus residue at $z_4=1+w$, then minus residue at $z_2=-1-2\ep-w$
and expand in $\ep$.
The resulting integral over $w,z_1$ and $z_5$ is denoted by $B_6$.
\item
Change the nature of the first pole of
$\Gm(-2 - 2 \ep - 2 w - z_2 - z_3)$, then take minus residue at
$z_4=-z_2-z_3$,
then a residue at $z_3=2\ep-z_2$, then take a residue at $z_2=2\ep$
and expand in $\ep$.
The resulting integral over $w,z_1$ and $z_5$ is denoted by $B_7$.
\end{enumerate}

One can see that all other contributions vanish at $\ep=0$.
By a suitable change of variables, one can observe that $B_7=B_6$.
In fact, the dependence of the
first five contributions on the Mandelstam variable $t$ is trivial:
they are just proportional to $1/t$.

The two-dimensional integrals $B_1$ and $B_3$ are products of one-dimensional
integrals which are
taken by closing contour to the left and summing up
resulting series with the help of formulae related to
those of Appendix~B.2 of \cite{DK}.
The three-dimensional integral $B_2$ is evaluated by closing the integration
contours over $z_1$ and $z_5$ to the left, summing up resulting series
and applying a similar procedure to a final integral in $z_2$.
The corresponding result is naturally expressed through
polylogarithms, up to Li$_3(x)$ depending on $s$ and $m^2$ in terms of
the variable
\[
v=\left[\frac{\sqrt{1-s/(4m^2)}+\sqrt{-s}/(2m)}
{\sqrt{1-s/(4m^2)}-\sqrt{-s}/(2m)}\right]^2\,.
\]

The form of this result provides a hint about possible functional
dependence of results for the three-(four-) dimensional integrals
$B_4$ ($B_5$), and a heuristic procedure which was explicitly formulated
in \cite{FKV} turns out to be successfully applicable here. First,
all the contributions, in particular
$B_4$ and $B_5$, are analytic functions of $s$ in a vicinity of the origin.
One can observe that any given term of the Taylor
expansion can be straightforwardly evaluated because the corresponding
integrals over $z_2$ and $z_4$ are recursively taken.
It is, therefore, possible to evaluate enough first terms (say, 30)
of this Taylor expansion.
Then one takes into account the type of the
functional dependence mentioned above,
turns to a new Taylor series in terms of the variable $v-1$
and assumes that
the $n$-th term of this Taylor series is a linear combination,
with unknown coefficients, of the following quantities:
\[
\frac{1}{n^4}, \;\; \frac{S_1(n)}{n^3}, \;\; \frac{S_2(n)}{n^2}, \;\;
\frac{S_1^2(n)}{n^2}, \;\; \frac{S_3(n)}{n}\ldots,
\]
where $S_k(n)=\sum_{j=1}^n j^{-k}$.
(Here some terms of the fourth order are listed.
See \cite{FKV} for more details.)
Using information about the first terms of the Taylor series
one solves a system of linear equations, finds those unknown coefficients
and checks this solution with the help of the next Taylor coefficients.

This experimental mathematics has turned out to be quite successful for the
evaluation of $B_4$ and $B_5$. Finally, the contribution $B_6$
is a product of a one-dimensional integral in $z_1$, which is easily
evaluated, and a two-dimensional integral in $w$ and $z_5$ which
involves a non-trivial dependence on $t$ and is evaluated
by closing the integration contours in $z_5$ to the left, summing up
a resulting series in terms of Gauss hypergeometric function
for which one can apply a parametric representation.
After that the internal integral over $w$ is taken by the same procedure
and, finally, one takes the parametric integral.

The final result takes the following form:
\be
B^{(0)}(s,t,m^2;\ep) =
-\frac{\left(i\pi^{d/2}
\e^{-\gm_{\rm E}\ep} \right)^2 x^2}{s^2 (-t)^{1+2\ep}}
\left[ \frac{b_2 (x)}{\ep^2} + \frac{b_1 (x)}{\ep}
  + b_{01} (x) + b_{02} (x,y)
+ O(\ep) \right]
\label{Result}
\;,
\ee
where $x=1/\sqrt{1-4m^2/s},\; y=1/\sqrt{1-4m^2/t}$, and
\bea
b_2 (x) &=&  2  (m_{x} - p_{x})^2\;,
\label{ResultEp2}
\\  
b_1 (x) &=&
 -8 \left[\Li{3}{ \frac{1 - x}{2}} + \Li{3}{\frac{1 + x}{2}} +
              \Li{3}{ \frac{-2 x}{1 - x}} + \Li{3}{\frac{2 x}{1 + x}}\right]
\nn \\ &&  
+ 4 (m_{x} - p_{x})\left[\Li{2}{\frac{1 - x}{2}}
- \Li{2}{\frac{-2 x}{1 - x}}\right] - (4/3) m_{x}^3 + 4 m_{x}^2 p_{x}
\nn \\ &&  
         - 6 m_{x} p_{x}^2 + (2/3) p_{x}^3
        + 4 l_{2} (m_{x} p_{x} + p_{x}^2)
- 2 l_{2}^2 (m_{x} + 3 p_{x})
\nn \\ &&  
- (\pi^2/3)(4 l_{2} - m_{x} - 3 p_{x}) + (8/3) l_{2}^3 + 14 \zeta_3
\;,
\label{ResultEp1}
\\
b_{01} (x) &=&
-8 (m_{x} - p_{x})\left[\Li{3}{ x} - \Li{3}{ -x} - \Li{3}{\frac{1 + x}{2}}
+ \Li{3}{\frac{1 - x}{2}}
\right.\nn \\ && \left.  \hspace*{-5mm}
- \Li{3}{\frac{2 x}{1 + x}}
+ \Li{3}{\frac{-2 x}{1 - x}}\right]
+ 4 \left[\Li{2}{ x}^2 + \Li{2}{ -x}^2 + 4 \Li{2}{\frac{1 - x}{2}}^2\right]
\nn \\ &&  
- 8  \Li{2}{ x}\Li{2}{ -x}
+ 16 \Li{2}{\frac{1 - x}{2}}(\Li{2}{ x} - \Li{2}{ -x})
\nn \\ &&  \hspace*{-15mm}
- (4/3) [\pi^2 - 6 l_{2}^2 + 3 m_{x}^2 + 6 m_{x} (2 l_{2} - 2 l_{x} - p_{x})
+ 12 l_{x} p_{x} - 3 p_{x}^2](\Li{2}{ x} - \Li{2}{ -x})
\nn \\ &&  
- (8/3) [\pi^2 - 6 l_{2}^2 + 6 l_{x} p_{x}
- 6 m_{x}( l_{x} + p_{x} - 2 l_{2})]\Li{2}{\frac{1 - x}{2}}
\nn \\ &&  
+ 8 (m_{x} - p_{x})\left[
(p_{x} - m_{x} + 2 l_{2}) \Li{2}{\frac{2 x}{1 + x}}
+ 2 (l_{x} - m_{x} + l_{2})\Li{2}{\frac{-2 x}{1 - x}}\right]
\nn \\ &&  
-8  (m_{x} - p_{x}) (2 l_{x} - p_{x} - 5 m_{x} + 4 l_{2})
(- m_{x} p_{x} + l_{2}( m_{x} + p_{x}) - l_{2}^2 + \pi^2/6)  
\nn \\ &&  
- (20/3)m_{x}^4 + (164/3)m_{x}^3 p_{x} - 40m_{x}^2 p_{x}^2
- (4/3) m_{x} p_{x}^3 - (8/3) p_{x}^4
\nn \\ &&  
+ 8m_{x} l_{x} (m_{x}^2 - 3m_{x} p_{x} + 2p_{x}^2)
\nn \\ &&  
- 4 l_{2} (7m_{x}^3 + 21m_{x}^2 p_{x} - 4m_{x} l_{x} p_{x}
- 23m_{x} p_{x}^2 + 4l_{x} p_{x}^2 - p_{x}^3)
\nn \\ &&  
- \pi^2 ((17/3)m_{x}^2 - (4/3)m_{x} l_{x} - 2m_{x} p_{x}
+ (4/3)l_{x} p_{x} - (7/3) p_{x}^2)
\nn \\ &&  
+ l_{2}^2 (84m_{x}^2 - 8m_{x} l_{x} - 16m_{x} p_{x}
+ 8l_{x} p_{x} - 44p_{x}^2)
\nn \\ &&  
- (8/3) l_{2} (6 l_{2}^2 - \pi^2) (3 m_{x} - 2 p_{x})
- (4/3)\pi^2 l_{2}^2 + 4l_{2}^4 + \pi^4/9
\;.
\label{ResultEp01}
\eea
The last piece of the finite part comes from $B_6$ and $B_7$:
\bea
b_{02} (x,y) &=&
2 ( p_{x} - m_{x})
\left\{
4\left[ \Li{3}{\frac{1 - x}{2}} - \Li{3}{\frac{1 + x}{2}}
+ \Li{3}{\frac{(1 - x) y}{1 - x y}}
\right.\right.
\nn \\ &&  \left.
- \Li{3}{\frac{-(1 + x) y}{1 - x y}} + \Li{3}{\frac{-(1 - x) y}{1 + x y}}
- \Li{3}{\frac{(1 + x) y}{1 + x y}} \right]
\nn \\ &&  
+ 2 \left[ \Li{3}{\frac{(1 + x)(1 - y)}{2(1 - x y)}}
- \Li{3}{\frac{(1 - x) (1 + y)}{2(1 - x y)}}
\right.\nn \\ &&  \left.
- \Li{3}{\frac{(1 - x) (1 - y)}{2(1 + x y)}}
+ \Li{3}{\frac{(1 + x) (1 + y)}{2(1 + x y)}} \right]
\nn \\ &&  \hspace*{-20mm}
+ 2 (m_{y} + p_{y} - m_{xy} - p_{xy})\left[2\Li{2}{ x} - 2\Li{2}{ -x}
+ \Li{2}{\frac{-2 x}{1 - x}} - \Li{2}{\frac{2 x}{1 + x}}\right]
 \nn \\ &&  
+ 4 (m_{xy} - p_{xy}) (\Li{2}{ -y} - \Li{2}{ y})
- 4 (m_{x} + p_{x} - 2l_{2}) \Li{2}{\frac{1 - x}{2}}
\nn \\ &&  
- 4 (m_{xy} - p_{xy}) \Li{2}{\frac{1 - y}{2}}
- 4 (m_{x} + l_{y} - m_{xy} )\Li{2}{\frac{(1 - x) y}{1 - x y}}
\nn \\ &&  \hspace*{-10mm}
+ 4(p_{x} + l_{y} - m_{xy})\Li{2}{\frac{ -(1 + x) y}{1 - x y}}
- 4 (m_{x} + l_{y} - p_{xy} )\Li{2}{\frac{-(1 - x) y}{1 + x y}}
\nn \\ &&  
+ 4 (p_{x} + l_{y} - p_{xy})\Li{2}{\frac{(1 + x) y}{1 + x y}}
\nn \\ &&  
+ 2 (m_{x} + p_{x} + m_{y} + p_{y} - 2m_{xy} - 2l_{2})
\Li{2}{\frac{(1 - x) (1 + y)}{2(1 - x y)}}
\nn \\ &&  
+ 2 (m_{x} + p_{x} + m_{y} + p_{y} - 2p_{xy} - 2l_{2})
\Li{2}{\frac{(1 - x) (1 - y)}{2(1 + x y)}}
\nn \\ &&  
+    
2 p_{x}^2(m_{y} + p_{y} - m_{xy} - p_{xy})
+ 2 p_{x} (2 (m_{y} l_{y} + m_{y} p_{y} + l_{y} p_{y})
\nn \\ &&  
+ m_{xy} (-m_{y} - 2l_{y} - 3p_{y} + 3m_{xy})
+ p_{xy}(-3m_{y} - 2l_{y} - p_{y} + 3p_{xy}))
\nn \\ &&  
+ 2 m_{x} (2p_{x} + m_{y} - 2l_{y} + p_{y})
(m_{y} + p_{y} - m_{xy} - p_{xy})
- p_{y}^2(m_{xy} + p_{xy})
\nn \\ &&  
+ 2p_{y}(2m_{xy}^2 + p_{xy}^2)
+ m_{y}^2(2p_{y} - m_{xy} - p_{xy})
\nn \\ &&  
+ 2m_{y}(p_{y}^2 + m_{xy}^2 + 2p_{xy}^2 - p_{y}(3m_{xy} + p_{xy}))
- 2(m_{xy}^3 + p_{xy}^3)
\nn \\ &&  
+ 2 l_{2}(
(4m_{y} + 4p_{y} - 3m_{xy})m_{xy} + (2m_{y} + 2p_{y} - 3p_{xy})p_{xy}
\nn \\ &&  
- 2(p_{x} + 2m_{x})
(m_{y} + p_{y} - m_{xy} - p_{xy}) - m_{y}^2 - 4m_{y} p_{y} - p_{y}^2)
\nn \\ &&   
+ 2l_{2}^2(3(m_{y} + p_{y}) - 2(2m_{xy} + p_{xy}))
\nn \\ && \left. 
- (\pi^2/3) (m_{y} + p_{y} - 8m_{xy} + 6p_{xy})
\right\}
\;.
\label{ResultEp02}
\eea
Here  $\Li{a}{z}$ is a polylogarithm \cite{Lewin}.
The following abbreviations are also used: $\zeta_3=\zeta(3)$,
$l_z =\ln z$ for $z=x,y,2$,  $p_z =\ln(1+z)$ and
$m_z =\ln(1-z)$ for $z=x,y,xy$.

This result is presented in such a way that it is manifestly real
at small negative values of $s$ and $t$. From this Euclidean domain,
it can be easily analytically continued to any other domain.

{\bf 3.}
The result~(\ref{Result})--(\ref{ResultEp02})
is in agreement with the leading
power behaviour in the (Sudakov) limit
of the fixed-angle scattering, $m^2 \ll |s|, |t|$.
This asymptotics is obtained by use of the strategy of
expansion by regions  \cite{BS,SR,Sb}.
The structure of regions is very rich.
The following family of seventeen regions participates here:
\vspace{1mm}

(h--h), (1c--h), \ldots, (4c--h), (1c--1c), \ldots, (4c--4c),

(1c--3c), (2c--4c), (1c--4c), (2c--3c),

(1uc--2c), (2uc--1c), (3uc--4c), (4uc--3c).
\vspace{1mm}

\noindent
Here $h$ denotes hard, $c$ -- collinear and $uc$ -- ultracollinear
regions for the two loop momenta.
(See \cite{SR} and Chapter~8 of \cite{Sb} for definitions of these regions.)
In particular, the (h--h) contribution is nothing but the massless
on-shell double box \cite{K1}.
The contributions (1c--1c), (3c--3c), (1c--3c) as well as the symmetrical
contributions   (2c--2c), (4c--4c) and (2c--4c)
are not individually regularized by dimensional regularization.
The poles in the auxiliary analytic regularization turn out to be
of the second order and are cancelled in the sum. After adding the
rest of the contributions, the poles of the third and fourth order
in $\ep$ are cancelled. Following this procedure, we obtain
\bea
B^{(0)}(s,t,m^2;\ep) &=&
-\frac{\left(i\pi^{d/2}
\e^{-\gm_{\rm E}\ep} \right)^2}{s^2 (-t)^{1+2\ep}}
\left\{ 2 \frac{L^2}{\ep^2}
-\left[ (2/3) L^3+(\pi^2/3) L+2\zeta_3\right]
\frac{1}{\ep}
\right.
\nn \\ &&  
-(2/3)L^4 +2\ln(t/s) L^3
-2( \ln^2(t/s) +4\pi^2/3) L^2
\nn \\ &&  \hspace*{-37mm}
+\left[
4 \Li{3}{-t/s} -4 \ln(t/s) \Li{2}{-t/s}
+(2/3)  \ln^3(t/s)-2\ln(1+t/s)  \ln^2(t/s)
\right.\nn \\ &&  \left.\left. \hspace*{-37mm}
+(8\pi^2/3) \ln(t/s)-2\pi^2 \ln(1+t/s)+10\zeta_3
\right] L +\pi^4/36 \right\}
+ O(m^2 L^3,\ep)
\;,
\label{LO}
\eea
where $L=\ln(-m^2/s)$. This asymptotic behaviour is reproduced
when one starts from result
(\ref{Result})--(\ref{ResultEp02}).

The analytical result presented above agrees also with results based
on numerical integration in the space of alpha parameters \cite{BH}
(where the 1\% accuracy for the $1/\ep$ and $\ep^0$ parts is guaranteed).

Let us stress that, in the present case with a non-zero mass,
there are no collinear divergences and the poles in $\ep$ are only up to
the second order, so that the resolution of singularities in $\ep$
in the MB integrals is relatively simple. Therefore, it looks promising
to use the technique presented, starting from (\ref{6MB}),
for the evaluation of any given master integral.
The construction of a recursive algorithm that would express
any given planar double box through some family of master integrals
is one of the next problems, as well as similar problems for the
evaluation of massive on-shell non-planar double-box Feynman
integrals.
Another possible scenario in the situation, where the mass is small but
still not negligible, is to evaluate the leading power (and all logarithms)
asymptotics when $m\to 0$. Anyway, the (h-h) contribution to
any such asymptotic behaviour is obtained by the algorithms of
\cite{K1,SV,Tausk,AGO,ATT}.

It is interesting to note that, in the above result,
there are no so-called two-dimensional harmonic polylogarithms \cite{2dHPL}
which have turned out to be adequate functions to express results for
the double boxes with one leg off-shell \cite{GR2}.
It is also an open question whether this phenomenon takes place
for general massive on-shell double boxes.

\vspace{0.5 cm}

{\em Acknowledgments.}
I am grateful to B.~Kniehl for kind hospitality during my visit to
the University of Hamburg in
October--November 2001 where this work was completed.
I am thankful to T.~Binoth and G.~Heinrich for comparison of the
presented result with their results based on numerical integration.
Many thanks to A.I.~Davydychev and O.L.~Veretin for
assistance in summing up one-dimensional power series
and to Z.~Bern and L.~Dixon for useful discussions.
This work was supported by the Russian Foundation for Basic
Research through project 01-02-16171, by INTAS through grant 00-00313,
and by the Bundesministerium f\"ur Bildung und Forschung
through Grant No.\ 05~HT1GUA/4.

\end{document}